\documentclass[multphys,vecphys]{svmult}

\usepackage{makeidx}         
\usepackage{graphicx}        
\usepackage{multicol}        
\usepackage[bottom]{footmisc}

\makeindex             

\begin{document}

\title*{3D-Spectroscopy of extragalactic planetary nebulae as diagnostic probes for galaxy evolution}
\titlerunning{XPN Physics}
\author{Andreas Kelz \and
Ana Monreal-Ibero \and Martin M. Roth \and Christer Sandin \and \\
Detlef Sch\"onberner \and Matthias Steffen}
\authorrunning{Kelz, Monreal-Ibero, Roth et al.}
\institute{Astrophysikalisches Institut Potsdam, An der Sternwarte 16, D-14482 Potsdam, Germany
\texttt{akelz@aip.de}}
%
%
\maketitle


\section{Introduction}
\label{sec:1}
In addition to study extragalactic stellar populations in their integrated light, the detailed analysis of individual resolved objects has become feasible, mainly for luminous giant stars and for extragalactic planetary nebulae (XPNe) in nearby galaxies.
A recently started project at the Astrophysical Institute Potsdam (AIP),  called ``XPN--Physics'', aims to verify if XPNe are useful probes to measure the chemical abundances of their parent stellar population.
The project involves theoretical and observational work packages.
Amongst other techniques, the use of 3D-spectroscopy is applied, despite the fact that XPNe are point-like sources, as it allows an accurate recording and hence subtraction of the complex underlying background.

\section{XPN Physics}
\label{sec:2}

Due to their bright emission in [O III]$\lambda$5007 and H$\alpha$, extragalactic Planetary Nebulae have been found in large numbers in the local group and out to galaxies in the Virgo and Coma clusters. Also, XPN were detected in the intra-cluster space of Virgo \cite{feldmeier-iaus04} and Coma \cite{gerhard-apjl05}. As XPNe can be individually resolved, they may serve as tracers for the low to intermediate mass stellar populations \cite{roth-an04}\cite{walsh-spie00} that otherwise can be studied only in the integrated light of its unresolved stars.

\begin{figure}
\centering
\includegraphics[width=0.3\textwidth,angle=-90]{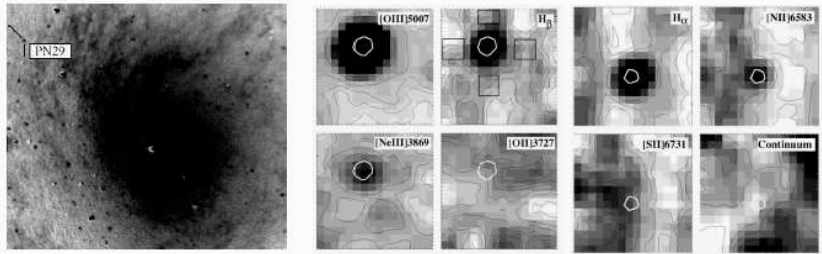}
\caption{Left: H$\alpha$  narrow-band image of the central 3 arcminutes of M31, obtained with Calar Alto 3.5 m prime focus camera+Fabry-P\'erot \cite{roth-apj04}.
The continuum has been subtracted using an off-band image to reveal the complex and fuzzy emission structure of the inter-stellar medium. The black dots mark XPN candidates, while spots with white substructures are indicative of residuals from imperfectly subtracted continuum sources (i.e. stars).
Right: Re-constructed monochromatic maps from PMAS \cite{roth-pasp05} observations of PN29 \cite{roth-apj04}, illustrating the problem of a highly variable background surface brightness distribution that changes with wavelength. 2D-sampling is crucial as otherwise the sky subtraction depends on slit orientation (as indicated in the H$\beta$ map). North is to the top, and east to the left.}
\label{kelz:fig1}
\end{figure}

{\bf Observational aspects:} For diagnostic purposes, spectroscopy of much weaker lines, such as [O III]$\lambda$4363 or [O II]$\lambda$3727, is needed. Because of the underlying, highly variable background which contaminates the spectra, 3D-spectroscopy has proven to be the preferred observational technique \cite{roth-eso06}. It avoids slit-losses, and the resulting data cubes offer the possibility to correct for effects caused by atmospheric dispersion \cite{kelz-an04}.
Furthermore, the 2-dimensional field-of-view of the integral-field-unit avoids the instrumental effects that are otherwise caused by slit orientation.
Figure~\ref{kelz:fig1} presents various re-constructed, monochromatic maps of a planetary nebula (PN29) in the Andromeda galaxy. Clearly, the crowded, unresolved background is not only
complex, but also varies with wavelength. By extracting various `slit'-spectra from the 3D-data cube, it could be shown that the resulting intensities for (weak) lines are dominated by the background subtraction \cite{roth-apj04}. The background coverage is a function of the slit orientation, even if the PN is always centered on the virtual slit.
However, using a deconvolution algorithm, the 3D data can be separated into a background and an object channel, thus yielding improved results in the determination of the line intensities.

\begin{figure}
\centering
\includegraphics[width=0.4\textwidth,angle=-90]{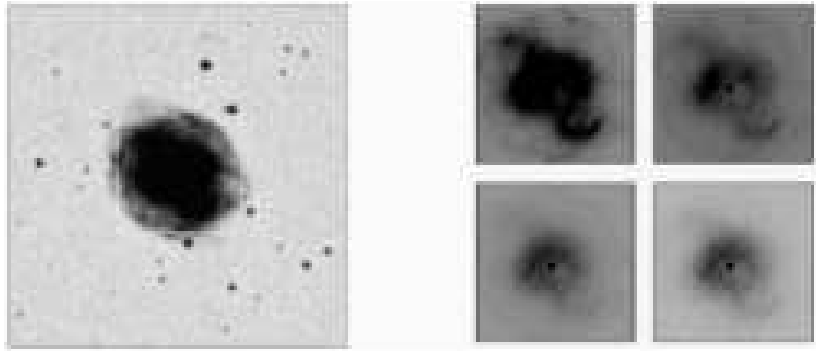}
\caption{NGC 4361 observed with the VIMOS-IFU at the VLT on April 18, 2004. Left: A Palomar image for comparison of the observed region (white square). Right: The panels show a 2x2 mosaic of four snapshot pointings as monochromatic maps obtained from the data cube in the wavelengths of [O III] 5007, [O III] 4959, H$\beta$, He II 4686 \cite{monreal-apjl05}. This object is a metal-poor PN from the halo-population of the Milky Way and will be used as a reference to asses the theoretical model calculations.}
\label{kelz:fig2}
\end{figure}

{\bf Theoretical aspects:} All aspects, the photoionization of the circumstellar
gas from the hot central star, the hot stellar wind and the occurrence of shock
waves, play an important role in shaping the ionized nebula \cite{perinotto04)}.
Conventional diagnostic analyses of PNe often neglect the time-dependence of
these important mechanisms, treating the nebula as a homogeneous and static
object in energy and ionization equilibrium. Within the ``XPN--Physics" project,
time-dependent hydrodynamical simulations \cite{schoenberner-iaus03} are used in
connection with selected observations, also of resolved galactic PNe (see Fig.\ref{kelz:fig2}) as to assess systematically, in how far common plasma diagnostic techniques are effected by density structure, dynamical evolution, and metalicities.

Using the combined approach of model simulations tested by specific observations,
the XPN-Physics programme aims to address topics such as: \\
-- How are structure and expansion of a PN determined by its metalicity? \\
-- What is the influence of the strong winds from central stars? \\
-- Does a minimum metalicity exist, necessary for the formation of a PN? \\
-- What are the quantitative differences if hydrodynamic simulations are used        for the plasma diagnostic instead of static photoionization codes? \\
-- Which ionization factors follow from hydrodynamic models?\\
-- Are XPNe useful probes of resolved stellar populations in galaxies?\\
\\
Beyond the research related to planetary nebulae in particular, it is expected that the results from this project will make a contribution to
the overall topic of spectral analysis of resolved stellar populations
which is one of the science cases for future ELTs.

%
%

{\bf Acknowledgements:} The XPN-Physics project is funded by the German Research Society DFG with grant number SCHO 394/26.




\printindex
\end{document}